\documentclass[a4paper,11pt]{article}
\usepackage[utf8]{inputenc}
\usepackage[T1]{fontenc}
\usepackage{mathcomp, textcomp}

\newcommand{\mmu}{\mbox{\sf \textmu}}

\pdfoutput=1
\usepackage{jheppub} \usepackage[T1]{fontenc} 
\title{Holographic metals at finite doping}
\author[a]{Laura Cruciani}
\author[a,b]{Nicol\'as E. Grandi}
\affiliation[a]{Departamento de F\'\i sica, Facultad de Ciencias Exactas, UNLP, \\ C.C. 67, CP1900 La Plata,  \\ Argentina.}
\affiliation[b]{Instituto de F\' \i sica de La Plata, Conicet,\\ C.C. 67, CP1900 La Plata,  \\ Argentina. }
\emailAdd{lauracruc@gmail.com}
\emailAdd{grandi@fisica.unlp.edu.ar}
\abstract{We construct the electron star solution to the model that was recently proposed by Kiritsis and Li in order to describe a holographic superconductor at finite doping. We do so by finding a map between the doped model and the standard undoped one. In this way, we are able to describe the holographic metallic phase at finite doping. In particular, we study the gauge field fluctuations and find the dependence of the electric conductivity on the doping parameter.

}
\begin{document} 
\maketitle
\newpage
\section{Introduction}
\label{sec:introduction}
Since the first proposal of a concrete example of an holographic setup, made by Maldacena in 1997  \cite{Maldacena_1999}, gauge/gravity duality has become a useful tool to probe the physics of quantum field theories at strong coupling, and to understand properties of quantum gravity in terms of a dual weakly coupled field theory \cite{Penedones_2016}. This is an extremely active area, where research is faced from two complementary approaches, known as top-down and bottom-up. The first one uses String Theory to carefully develop a duality between a well defined gravitational background and a perfectly identified boundary field theory. The second one starts by selecting the relevant operators of the boundary field theory and then identifying their dual fields in the bulk by their quantum numbers. 
 
A particularly interesting application of bottom-up approach is that of low dimensional quantum field theories at finite density, as a proxy for strongly coupled Condensed Matter systems, a research area presently known as ``AdS/CMT'' for Condensed Matter Theory \cite{Hartnoll_2009, zaanen2015holographic}. The interest of such approach is to study those condensed matter systems in which strong coupling is believed to play a role, examples of which are fractional quantum Hall effect \cite{RevModPhys.71.S298} and the class of superconducting materials with high critical temperature known as High $T_c$ superconductors \cite{Anderson_2013, zaanen2010modern}.

In this context, there were proposals for the holographic dual of a superconducting phase with a $s$-, $p$- or $d$-wave order parameter \cite{Hartnoll_2008, Hartnoll_2008dos, Gubser_2008, Chen_2010, Benini_2010}. Other phases were also explored, for example the antiferromagnetic phase \cite{Iqbal_2010}, and the inhomogeneous stripped and checkerboard phases \cite{Rozali_2013, Donos_2013, withers2013moduli, Ling_2014}. Of particular interest for the present paper is the metallic phase \cite{Lee_2009, faulkner2010black, Faulkner1043, Faulkner_2011s, _ubrovi__2009, Sachdev_2010, de_Boer_2010, Arsiwalla2011, Hartnoll_2011, Hartnoll_2011dos, Hartnoll_2011tres, Puletti_2011} known as ``holographic metals''. Such construction reproduces correctly part of the phenomenology of the metallic phase on High $T_c$ superconductors or ``strange metal''. In particular, it has a resistivity that behaves linearly as a function of the temperature, a feature that is reproduced in the holographic setup \cite{PhysRevD.86.124046}.

Generically, the dual field theory in a holographic setup is a conformal field theory. Then, a finite chemical potential sets a scale in terms of which the temperature and magnetic field dependence is measured. In order to include a doping axes, an additional scale is needed. This was implemented by Kiritsis and Li by means of the introduction of a second chemical potential, playing the role of the impurities \cite{Kiritsis_2016}. Such model results on a phase diagram that, in a region of parameters, is qualitatively very similar to that of High $T_c$ compounds. At low enough temperature, it presents an antiferromagnetic region at small doping, a superconducting dome at intermediate doping, and a Fermi liquid metallic phase at large doping. Above the superconducting dome, the Fermi liquid behaves as a strange metal. The strange metal phase flows into a normal Fermi liquid as the doping grows, reproducing the  behavior of the High $T_c$ superconductors \cite{Giordano_2018}. The phase diagram can be enriched by including frame fields to simulate an underlying lattice \cite{Baggioli_2016_2}, and a finite magnetic field \cite{Giordano_2018}. 

In the present paper we extend previous developements on the Kiritsis and Li model to include backreaction on the metallic phase. We do so by constructing its electron star solution and exploring its perturbations. 

\section{The holographic model for the doped strange metal}
\label{sec:model.construction}
In the present section we introduce our holographic model, starting with the undoped case, extending it with an additional gauge field, and then proposing a new set of variables that maps the extended undoped model  into a model suitable to describe a doped metal. 
\subsection{The undopped holographic metal and the electron star solution}
\label{sec:undoped}
The model for the undoped holographic metal we are intereseted in was first described in  \cite{Hartnoll_2011}. In this section we sketch the main results, the interested reader is referred to the original publication.
\paragraph{The model:} The bulk dynamics of the undoped holographic metal is defined according to the action 
\begin{equation}
S=S_{\sf Gravity}+S^+_{\sf Maxwell}+S_{\sf Matter} \,.
\label{eq:action.total}
\end{equation}
The term $S_{\sf Gravity}$ represents the Einistein-Hilbert gravitational action in $3+1$ dimensions, 
\begin{equation}
S_{\sf Gravity}=
\frac{1}{2\kappa^2}
\int d^4x \; \sqrt[]{-g}
\left(
R \!+\! \frac{6}{L^2} \right)\,,
\label{eq:action.gravity.maxwell}
\end{equation}
where  $L$ is the AdS radius, while the term $S_{\sf Maxwell}$ includes a  Maxwell field $A_\mu^+$
\begin{equation}
S_{\sf Maxwell}^+=
-\frac{1}{4e ^2}
\int d^4x \; \sqrt[]{-g}\,
F_+^2\,.
\label{eq:action.gravity.maxwell.plus}
\end{equation}
For this Maxwell field the holographic interpretation is standard, its boundary value being coupled to a conserved particle current. On the other hand, the matter part is treated as a perfect fluid, for which we use a Schutz action modified to include electromagnetic coupling to the Maxwell field in the way proposed in  \cite{Hartnoll_2011}. It reads
\begin{equation}
S_{\sf Matter}=
\int d^4x \; \sqrt[]{-g}
\left( 
-\rho+\sigma u^\mu\left(\partial_\mu\phi+q_+\!A^+_\mu
\right)
+\lambda\left(u_\mu u^\mu+1\right)
\right)\,,
\label{eq:action.fluid}
\end{equation}
where $\rho$ is the fluid energy density considered as a function of  the fluid particle density $\sigma$, $u^\mu$ is the fluid four-velocity, the scalar field $\phi$ is the Clebsh potential, and $\lambda$ is a Lagrange multiplier ensuring a unitary time-like four-velocity. The charge of the matter particles is represented by $q_+$. The matter dynamics obtained from this action is the one used in \cite{Hartnoll_2011}, to represent the fermionic degrees of freedom of the boundary theory. 

To write an equation of state, that implements the functional relation between $\rho$ and $\sigma$, we define the local chemical potential of the fluid in the bulk as
$\mmu={d\rho}/{d\sigma}$. Relying on the Thomas-Fermi approximation, in which there is a large number of particles within a region of the size of an AdS radius. For fermions at zero temperature, we get
\begin{eqnarray}
\rho&=&\gamma\int_m^{{\mbox{\small {\sf \textmu} \normalsize}}}dE\,E^2\sqrt{E^2-m^2}\,,
\label{eq:rho}
\\
\sigma&=&\gamma\int_m^{{\mbox{\small {\sf \textmu} \normalsize}}}dE\,E \sqrt{E^2-m^2}\,,
\label{eq:sigma}
\end{eqnarray}
where $\gamma$ is a parameter related to the number of particle species.

\paragraph{The electron star:} The ``electron star'' configuration is a static and planar solution of the equations of motion obtained from \eqref{eq:action.total}-\eqref{eq:sigma}. 
It is found with the Ansatz  
\begin{eqnarray}
ds^2&=& L^2
\left(
-f dt^2 + g dz^2+\frac{1}{z^2}(dx^2+dy^2)
\right)\,,
\label{eq:ansatz.metric}
\\
A^+&=&\frac{e L}{\kappa}h^+\,dt\,,
\label{eq:ansatz.maxwel.plus}
\\
\phi&=&0\,,
\label{eq:ansatz.clebsh}
\\
u^\mu&=&\delta^{\mu0}\frac{1}{L\sqrt{f}}\,,
\label{eq:ansatz.velocity}
\end{eqnarray}
where the functions $f$, $g$ and $h^+$ are $z$-dependent.
The solution satisfies asymptotically AdS boundary conditions in the UV, with
\begin{eqnarray}
&&A_0|_{\sf boundary}^+ = \mu^+\,.
\label{eq:boundary.plus}
\end{eqnarray}
Here $\mu^+$ defines the chemical potential of the boundary conserved particles dual to the gauge field $A^+$. On the other hand, in the deep infrared $z\to\infty$ it has a Lifshitz form
\begin{eqnarray}
f=\frac{1}{z^{2{\sf z}}}\,,
\qquad\qquad
g=\frac{g_{\infty}}{z^2}\,,
\qquad\qquad
h^+=\frac{h^+_{\infty}}{z^{\sf z}}\,,
\label{eq:lifshitz}
\end{eqnarray}
with a dynamical critical exponent ${\sf z}$ and the constants $h_\infty$ and $g_\infty$ determined in terms of the number of local degrees of freedom $\gamma$ and the fermion mass $m$ and charge $q_+$. As we flow to the ultraviolet, the solution deforms and the local chemical potential $\mmu$ decreases until it reaches $m$ at the boundary of the star. Further to the AdS boundary, the solution matches into a planar Reissner-Nordstrom metric with mass $M$ and carge $Q^+$.

The grand canonical potential is obtained by evaluating on-shell the action \eqref{eq:action.total} and adding the necessary holographic renormalization counterterms. It reads
\begin{equation}
\Omega = M - \mu^+ Q^+\,.
\label{eq:grand.canonical.undoped}
\end{equation}

\paragraph{Fluctuations and conductivity:} By perturbing of the electron star, we can obtain the electrical conductivity of the boundary theory.
The vector fluctuations of the above electron star background are parameterized as
\begin{eqnarray}
&&A_x^+=\frac{eL}{\kappa}\delta A^+_x e^{-i\omega t}\,,
\label{eq:perturbation.gauge}
\qquad\quad
g_{tx}=L^2\,\delta g_{tx}e^{-i\omega t}\,,
\qquad\quad
u_x=L\,\delta u_xe^{-i\omega t}\,.
\end{eqnarray}
The resulting linearized equations of motion result in the UV behavior
\begin{equation}
\delta A^+_x\simeq \delta A_x^{\!+\,(0)}+
z\,\delta A_x^{\!+\,(1)}\,.
\label{eq:boundary.fluctuation}
\end{equation}
Imposing regular boundary conditions in the IR, the coefficients of the leading and subleading parts of \eqref{eq:boundary.fluctuation} become mutually dependent. In terms of them we can write the boundary conductivity
\begin{equation}
\sigma_{xx}^+=-\frac{i\,c}\omega\, \frac{\delta A_x^{\!+\,(1)}}{\delta A_x^{\!+\,(0)}}\,,
\label{eq:sigma.plus}
\end{equation}
where $c=\lim_{z\to 0}z^2f$.

Since the conductivity is dimensionless in $2+1$ dimensions, it scales as a constant at large frequencies \cite{Hartnoll_2009}. Conversely, in the limit of small frequencies its real part vanishes as $\sim \omega^2$, and the imaginary part has a pole \cite{Hartnoll_2011}. 

\subsection{The trivially extended holographic metal}
\label{sec:extended}
\paragraph{The model:} We extend the holographic model \eqref{eq:action.total} by adding a second Maxwell field $A^-_\mu$ with a standard action
\begin{equation}
S_{\sf Maxwell}^-=
-\frac{1}{4e^2}
\int d^4x \; \sqrt[]{-g}\,
F_-^2\,.
\label{eq:action.gravity.maxwell.minus}
\end{equation}
From the boundary perspective this new field is dual to a second conserved current of neutral particles. 
In this way, we obtain a trivially extended model 
\begin{equation}
S^\pm=S_{\sf Gravity}+S^+_{\sf Maxwell}+S^-_{\sf Maxwell}+S_{\sf Matter} \,.
\label{eq:action.total.plusminus}
\end{equation}
The matter part of the dynamics is unmodified, implying in particular that the bulk fluid is neutral with respect to the new Maxwell field $A^-_\mu$. The only coupling of $A_\mu^-$ with the remaining fields is then through its contribution to the energy momentum tensor.

\paragraph{The electron star:} To obtain an electron star solution of the trivially extended model, we supplement \eqref{eq:ansatz.metric}-\eqref{eq:ansatz.velocity} with an Ansats for $A_\mu^-$ with the form
\begin{equation}
A^-= \frac{eL}{\kappa}\,h^-\,dt\,.
\label{eq:ansatz.maxwel.minus}
\end{equation}
And with the corresponding boundary condition
\begin{eqnarray}
&&A_0|_{\sf boundary}^- = \mu^-\,.
\label{eq:mu.minus}
\end{eqnarray}
Since the Maxwell equations for $A_\mu^-$ have no sources, a consistent solution is the constant $h^-=(\kappa/e L) \mu^-$. The resulting electromagnetic curvature $F_{\mu\nu}^-$ vanishes, thus the second Maxwell field makes no contribution to the energy momentum tensor on the Einstein equations. In consequence, the electron star solution of the previous subsection is still a solution of the trivially extended model. The grand canonical potential keeps the form \eqref{eq:grand.canonical.undoped}.
\paragraph{Fluctuations and conductivities:} Now we have an additional conserved current in the boundary, whose conductivity we want to compute. To do that, we supplement \eqref{eq:perturbation.gauge}
with a perturbation for $A_\mu^-$ as
\begin{equation}
A_x^-=\delta A_x^-\,.
\label{eq:perturbation.gauge.minus}
\end{equation}
Since the unperturbed gauge curvature $F_{\mu\nu}^-$ vanishes, the contribution of $\delta A_x^-$ to the energy momentum tensor is quadratic. Then, to linear order the equations for $\delta A^+_x$, $\delta g_{tx}$ and $\delta u_x$ are unmodified, and the resulting $\sigma_{xx}^+$ is unchanged. Moreover, we have a new conductivity associated to the second boundary current, that we can write 
\begin{equation}
\sigma_{xx}^-=-\frac{i\,c}\omega\, \frac{\delta A_x^{\!-\,(1)}}{\delta A_x^{\!-\,(0)}}\,.
\label{eq:sigma.minus}
\end{equation}
This conductivity is independent of the frequency $\omega$ (see section \ref{sec:doped}). Now we also have 
\begin{equation}
\sigma_{xx}^\pm=-\frac{i\,c}\omega\, \frac{\delta A_x^{\!+\,(1)}}{\delta A_x^{\!-\,(0)}}\,,
\qquad\qquad
\sigma_{xx}^\mp=-\frac{i\,c}\omega\, \frac{\delta A_x^{\!-\,(1)}}{\delta A_x^{\!+\,(0)}}\,.
\label{eq:cross.conductivities}
\end{equation}
Since $\delta A_x^-$ is completely decoupled, a modification on its leading term $\delta A_x^{\!-\,(0)}$ does not affect the subleading term of the other field $\delta A_x^{\!+\,(1)}$, implying that $\sigma^\pm=0$. The same is true in the opposite direction, implying $\sigma^\mp=0$.

\subsection{The holographic model for the doped strange metal}
\label{sec:doped}
\paragraph{The model:} The above trivially extended model can be rewritten in a different form that allows a richer physical interpretation, defining a new pair of Maxwell fields as
\begin{eqnarray}
q_+A_\mu^+&=&qA_\mu+\tilde{q}\tilde{A}_\mu\,,
\qquad\qquad
q_+A_\mu^-=
\tilde{q}A_\mu-q\tilde{A}_\mu\,,
\label{eq:Aminus}
\end{eqnarray}
where the new charges $q$ and $\tilde{q}$ are chosen in the circle $q_+^2=q^2+\tilde{q}^2$. The action becomes
\begin{equation}
S=S_{\sf Gravity}+S_{\sf Maxwell}+\tilde{S}_{\sf Maxwell}+S_{\sf Matter} \,.
\label{eq:action.total.AAtilde}
\end{equation}
where $S_{\sf Gravity}$ is unmodified, while the fields $A_\mu$ and $\tilde{A}_\mu$ have standard Maxwell actions
\begin{eqnarray}
&&S_{\sf Maxwell}=
-\frac{1}{4e^2}
\int d^4x \; \sqrt[]{-g}\,
F^2\,,
\qquad\qquad
\tilde{S}_{\sf Maxwell}=
-\frac{1}{4e^2}
\int d^4x \; \sqrt[]{-g}\,
\tilde{F}^2\,.
\label{eq:action.gravity.maxwell.AAtilde}
\end{eqnarray}
The matter part on the other hand now corresponds to a fluid coupled to both Maxwell fields $A_\mu$ and $\tilde{A}_\mu$ as
\begin{equation}
S_{\sf Matter}=
\int d^4x \; \sqrt[]{-g}
\left( 
-\rho+\sigma u^\mu\left(\partial_\mu\phi+qA_\mu+\tilde{q}\tilde A_\mu\right)
+\lambda\left(u_\mu u^\mu+1\right)
\right)\,.
\label{eq:action.fluid.AAtilde}
\end{equation}

The model \eqref{eq:action.total.AAtilde} can be considered as a truncation of the model introduced in \cite{Kiritsis_2016} including only the gravitational and $U(1)\times U(1)$ gauge degrees of freedom. The holographic interpretation is standard for one of the $U(1)$ fields, say $A_\mu$, its boundary value being coupled to the conserved particle current for the charge carriers. The second $U(1)$ field, in this case $\tilde{A}_\mu$, stands for a second species of conserved particles, that can be loosely interpreted as the impurities. In practice, its boundary value sets a scale on the field theory that allows for the definition of a doping parameter (see bellow). 

\paragraph{The electron star:} applying the change of variables to the gauge fields of the trivially extended electron star Ansatz \eqref{eq:ansatz.maxwel.plus} and \eqref{eq:ansatz.maxwel.minus} and taking into account the constant solution for $h^-$, we immediately obtain an electron star solution for the present model, with 
\begin{eqnarray}
A&=&\frac{1}{q_+}\left(\frac{q e L}{\kappa}h^++\tilde{q}\,\mu^-\right)\,dt\,,
\qquad\qquad
\tilde{A}=
\frac{1}{q_+}\left(
\frac{\tilde{q}e L}{\kappa}h^+-q\,\mu^-\right)\,dt\,.
\label{eq:ansatz.maxwel.Atilde}
\end{eqnarray}
These fields have boundary values related to the chemical potential for the charge carrier and impurity currents respectively, that satisfy
\begin{eqnarray}
\mu&=&\frac{1}{q_+}\left(q\,\mu^++\tilde{q}\,\mu^-\right)\,,
\qquad\qquad
\tilde{\mu}=
\frac{1}{q_+}\left(
\tilde{q}\,\mu^+-q\,\mu^-\right)\,.
\label{eq:mutilde}
\end{eqnarray}

Notice that in   \eqref{eq:action.total.AAtilde} there is a complete symmetry between $A_\mu$ and $\tilde{A}_\mu$. This is manifest here since the change $q\to\tilde{q}$ and $\mu^-\to-\mu^-$ results in the exchange of $A_\mu$ with $\tilde{A}_\mu$.
Thus in what follows we choose one of them, say $\tilde{A}_\mu$, and use it to fix the scale trough its boundary value $\tilde{\mu}$. 
This allows us to define a doping parameter as
\begin{eqnarray}
{\sf x}&=&
\frac{\mu}{\tilde{\mu}}=
\frac
{
q\,\mu^++\tilde{q}\,\mu^-
}
{
\tilde{q}\,\mu^+-q\,\mu^-
}
\,.
\label{eq:doping}
\end{eqnarray}  

The grand canonical potential in terms of the new variables reads
\begin{equation}
\Omega = M - \mu \left(q + \frac{\tilde{q}}{\sf x}\right)\frac{Q^+}{q^+}\,.
\label{eq:grand.canonical}
\end{equation}
This depends on ${\sf x}$ explicitly, and also implicitly through $M$ and $Q^+$. 

\paragraph{Fluctuations and conductivities:}
with the new set of variables, we can define the conductivities
associated to the charge carriers and to the impurities as follows: if $J$ is the boundary current associated to $A$, and $\tilde{J}$ is the one associated to $\tilde{A}$, then we have
\begin{equation}
J = \frac{1}{q^+}\left(qJ^++\tilde{q}J^-\right)
\,,
\qquad\qquad\quad
\tilde{J} = \frac{1}{q^+}\left(\tilde{q}J^-+qJ^-\right)\,.
\label{eq:currents}
\end{equation}
By Kubo formula, we know that $\sigma\propto \langle J J\rangle$ and $\tilde{\sigma}\propto \langle\tilde{J}\tilde{J}\rangle$. Using the linear transformation \eqref{eq:currents} and the fact that $\langle J^+J^-\rangle=0$, we obtain 
\begin{eqnarray}
&&\sigma=
\frac{1}{q_+^2}\left(
q^2{\sigma^+}+\tilde{q}^2\sigma^-
\right)
\,,
\qquad\qquad\quad
\tilde{\sigma}=
\frac{1}{q_+^2}\left(
\tilde{q}^2{\sigma^+}+q^2\sigma^-
\right)\,.
\label{eq:conductivitytilde}
\end{eqnarray}
These conductivities are functions of the doping ${\sf x}$. They are interchanged when $q$ is interchanged with $\tilde{q}$.

In the expressions \eqref{eq:conductivitytilde}, the conductivity $\sigma^-$ is completely independent on the frequency $\omega$. This can be understood from the fact that the resulting linearized equations of motion for ${\delta A_x^-}$ at any frequency coincide with the limit of large frequencies of the corresponding equations for ${\delta A_x^+}$. Since we know that $\delta A_x^+$ at large frequencies is such that $\sigma^+$ is independent of $\omega$ \cite{Hartnoll_2009}, the same must be true at any frequency for ${\delta A_x^-}$ implying that $\sigma^-$ is also independent of $\omega$.

The form of the conductivity $\sigma$ as a function of the frequency $\omega$ can be obtained from the above referred behaviours of $\sigma^+$ and $\sigma^-$. At low frequency the real part of $\sigma^+$ vanishes as $\omega^2$  \cite{Hartnoll_2011}, implying that ${\rm Re}[\sigma]\sim (\tilde{q}^2/q^2_+){\rm Re}[\sigma^-]\equiv \sigma_0$, where $\sigma^-$ is a constant. Its imaginary part on the other hand has a pole, originated in the pole of ${\rm Im}[\sigma^{+}]$  \cite{Hartnoll_2009}. At large frequencies $\sigma^+$ goes to a real constant, thus the same is true for $\sigma\sim (\tilde{q}^2{\sigma^+}+q^2\sigma^-)/{q_+^2}\equiv \sigma_1$. In other words, the role of $q$ and $\tilde{q}$ is to shift the value of the real part of the conductivity at zero frequency, and its asymptotic limit.

\section{Results and discussion}
\label{sec:results}
\begin{figure}[ht]
\,
\includegraphics[width=.44\textwidth]{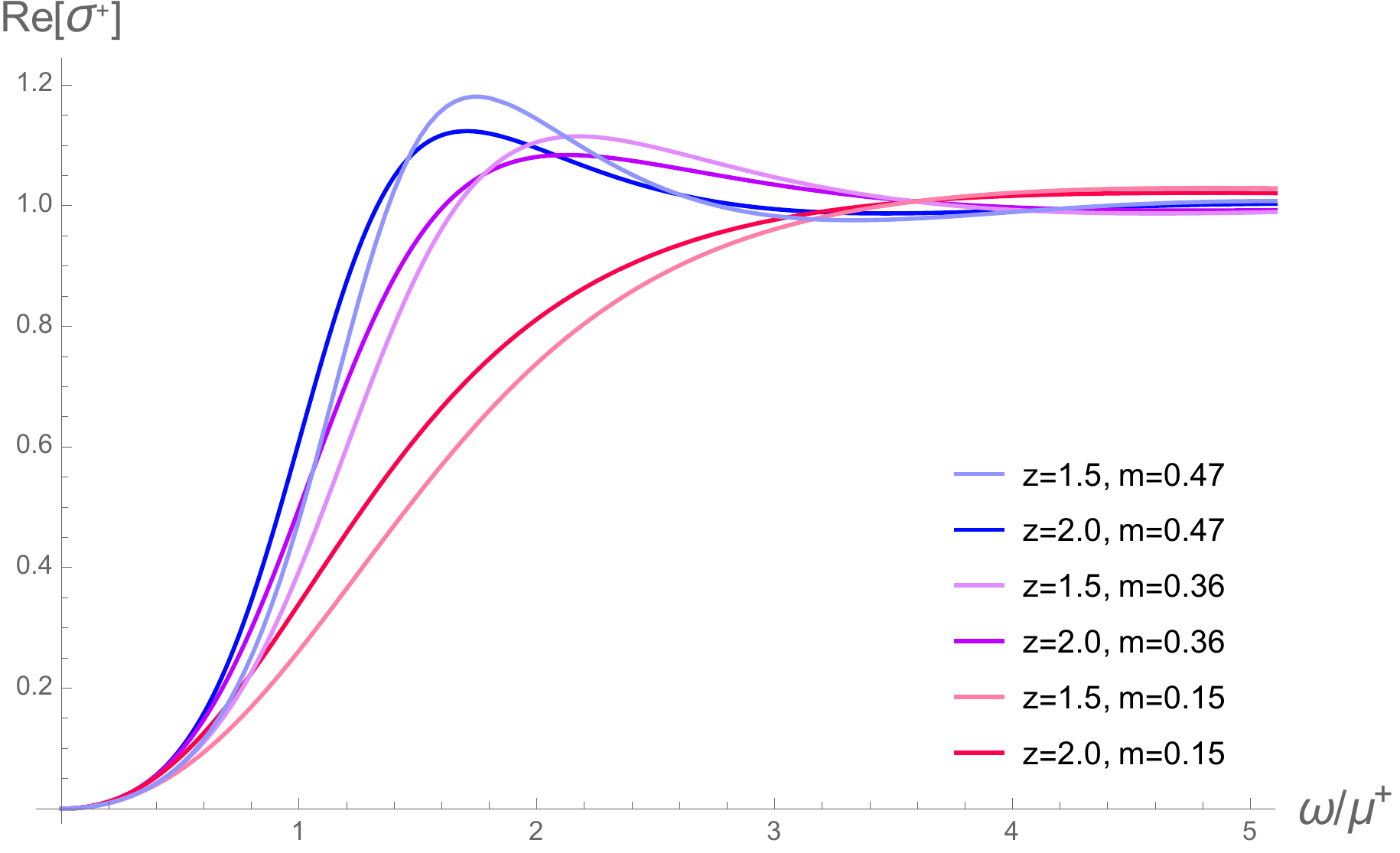}
\qquad\ \
\includegraphics[width=.44\textwidth]{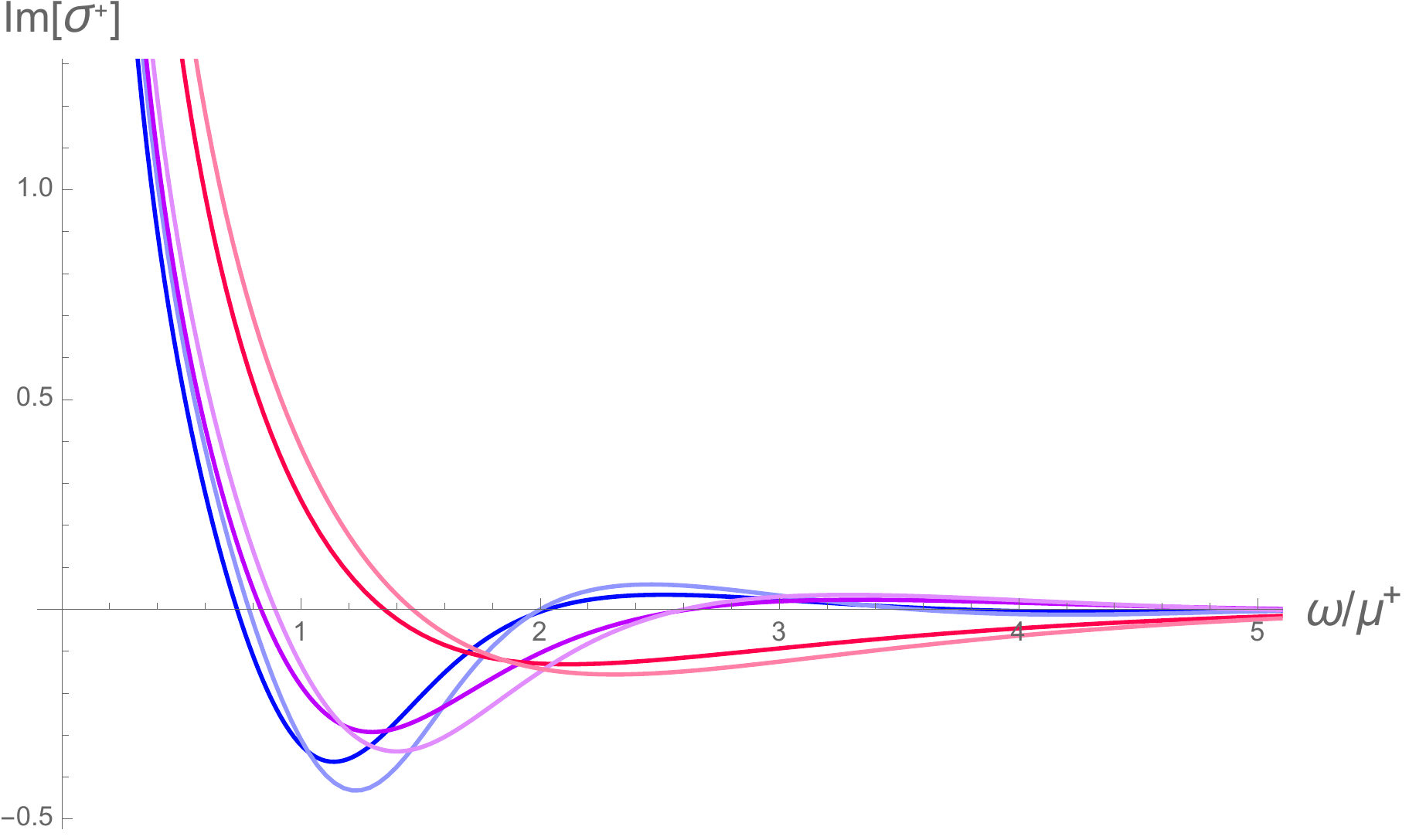}
\caption{\label{fig:figura0}
Undoped conductivity $\sigma^+$ as a function of the ratio $\omega/\mu^+$, for different values of the mass $m$ and the critical exponent ${\sf z}$ for $q^+=1$.
}
\end{figure}

We produced a family of electron star solutions of the undoped model \eqref{eq:action.total} with fixed mass $m$ and charge $q_+$ indexed by the value of the dynamical IR critical exponent ${\sf z}$. We calculated the corresponding conductivities $\sigma^+$ as in \eqref{eq:sigma.plus} as functions of the frequency $\omega$ reproducing the results of \cite{Hartnoll_2011}, see Fig. \ref{fig:figura0}. The real part vanishes at vanishing frequency, and flows at large frequencies into a constant value independent of ${\sf z}$. At lower values of ${\rm z}$, the real part of the conductivity has a peak at intermediate frequencies that gets diluted as ${\rm z}$ grows, the curve becoming flatter. Regarding the imaginary part, it diverges at zero frequencies, and goes to zero at large frequencies. For small values of ${\sf z}$, the imaginary part has a dip, that becomes smoother at larger ${\sf z}$. As the mass is increased, the peak on the real part of the conductivity, and the dip of the imaginary part, become more pronounced and shift towards smaller frequencies.

Then, using 
\eqref{eq:ansatz.maxwel.Atilde} with different values of chemical potential $\mu^-$ and the charge $q$, we mapped such family into different families of solutions of the doped model \eqref{eq:action.total.AAtilde}. Inside each of them, the value of the doping $\sf x$ defined as in equation \eqref{eq:doping} can be used as an index to label the solutions. 

\begin{figure}[ht]
\,
\includegraphics[width=.44\textwidth]{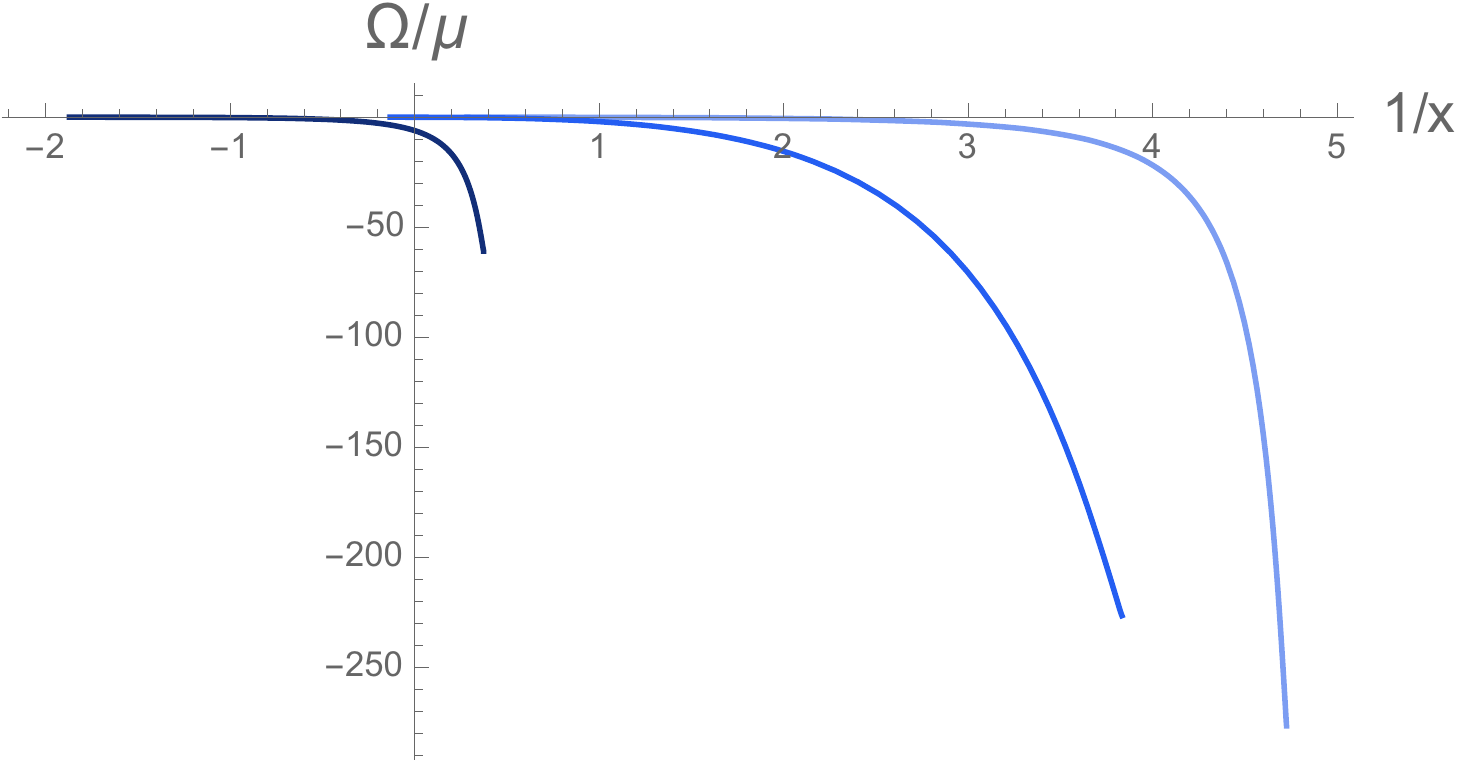}
\qquad\ \
\includegraphics[width=.44\textwidth]{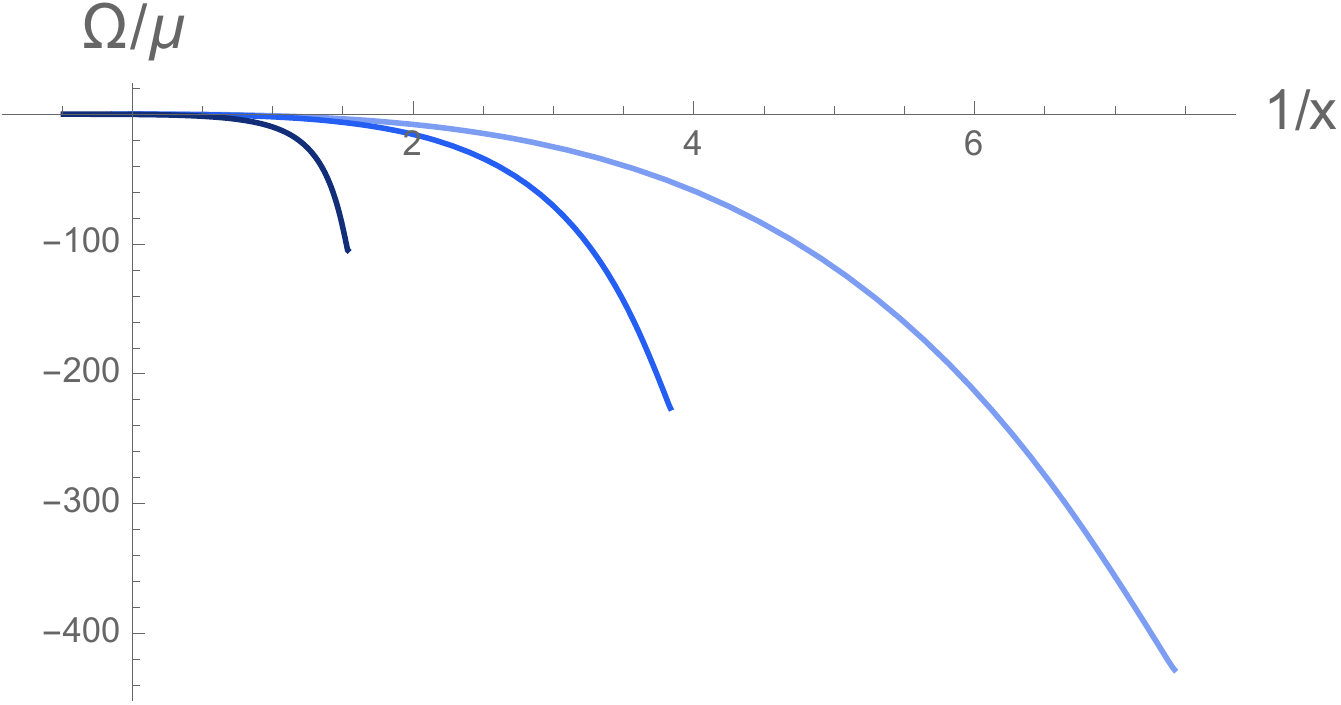}
\put(-257,-3){\tiny $\mu^-\!=\!0.2$\normalsize}
\put(-277,12){\tiny $\mu^-\!=\!1.5$\normalsize}
\put(-355,59){\tiny $\mu^-\!=\!2.8$\normalsize}
\put(-29,-3){\tiny $q\!=\!0.08$\normalsize}
\put(-100,35){\tiny $q\!=\!0.2$\normalsize}
\put(-147,58){\tiny $q\!=\!0.5$\normalsize}
\caption{\label{fig:figura4} Grand canonical potential as a function of the doping for different vales of the chemical potential $\mu^-$ (left) and of the charge $q$ (right).}
\end{figure}

\begin{figure}[h]
\centering
\vspace{1.2cm}
\includegraphics[width=.44\textwidth]{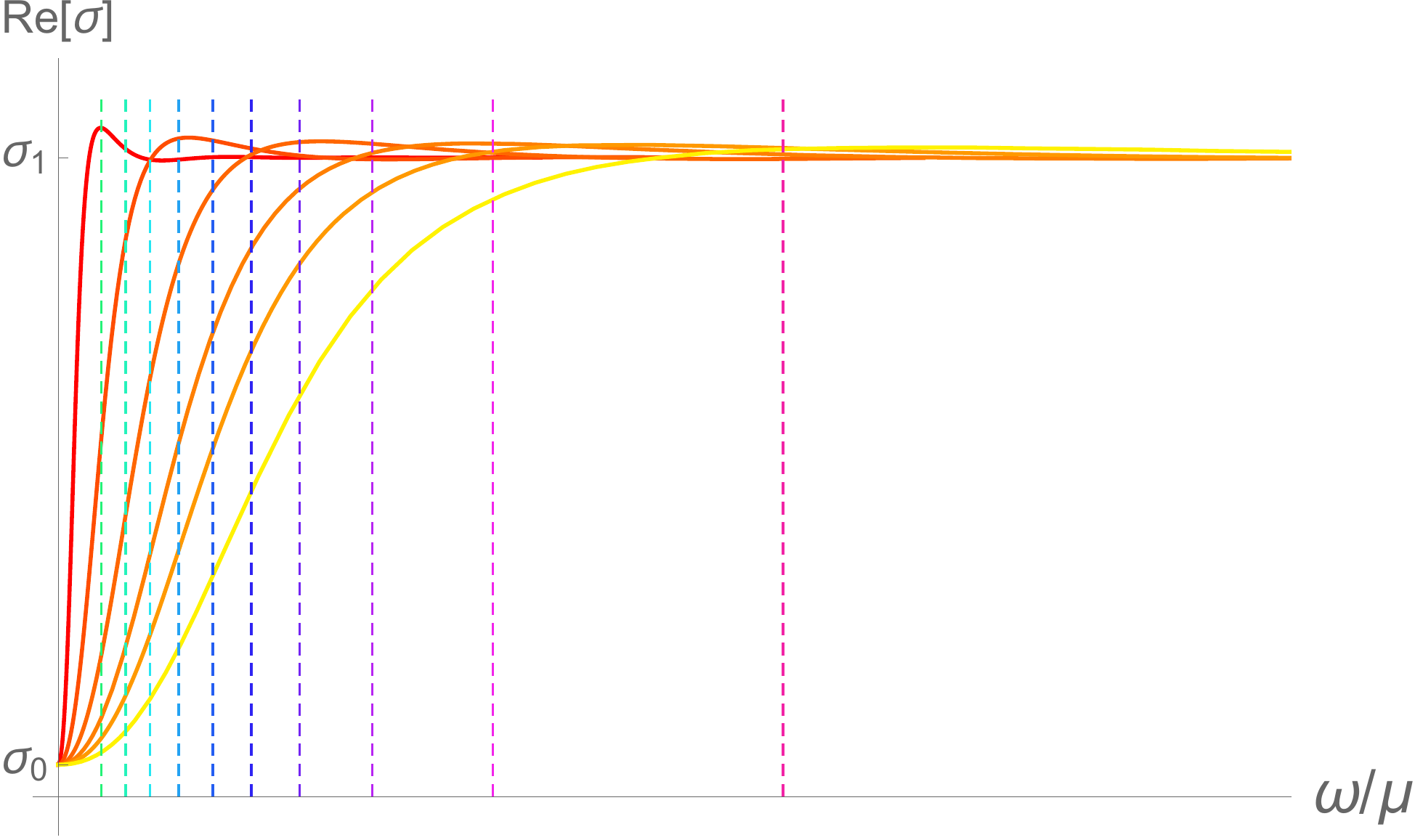}
\includegraphics[width=.07\textwidth]{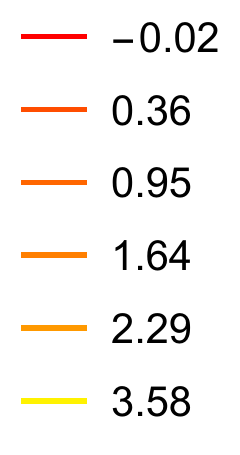}
\includegraphics[width=.44\textwidth]{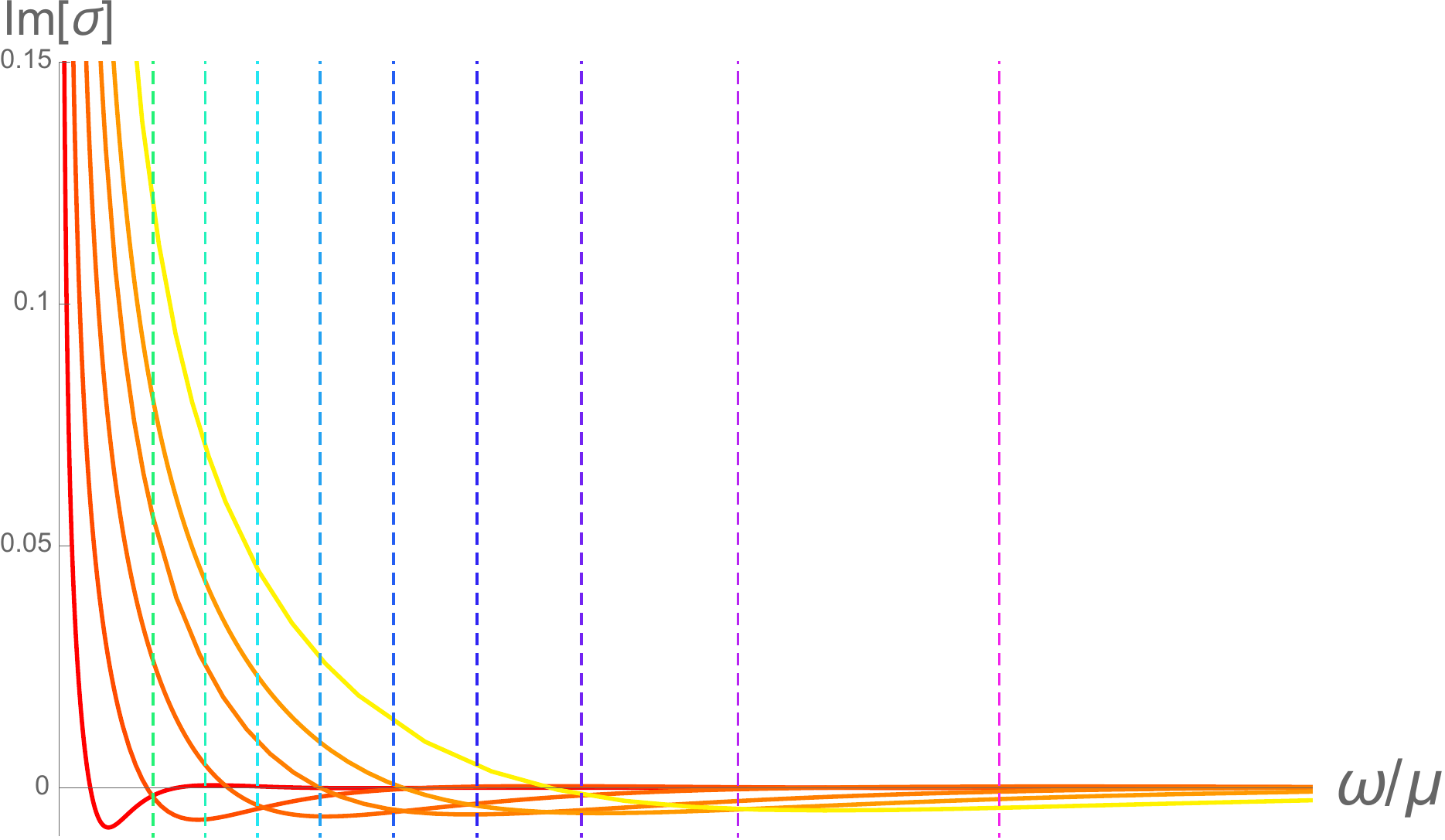}
\put(-215,62){\scriptsize $1/{\sf x}$\normalsize}
\\
\vspace{.4cm}
\includegraphics[width=.44\textwidth]{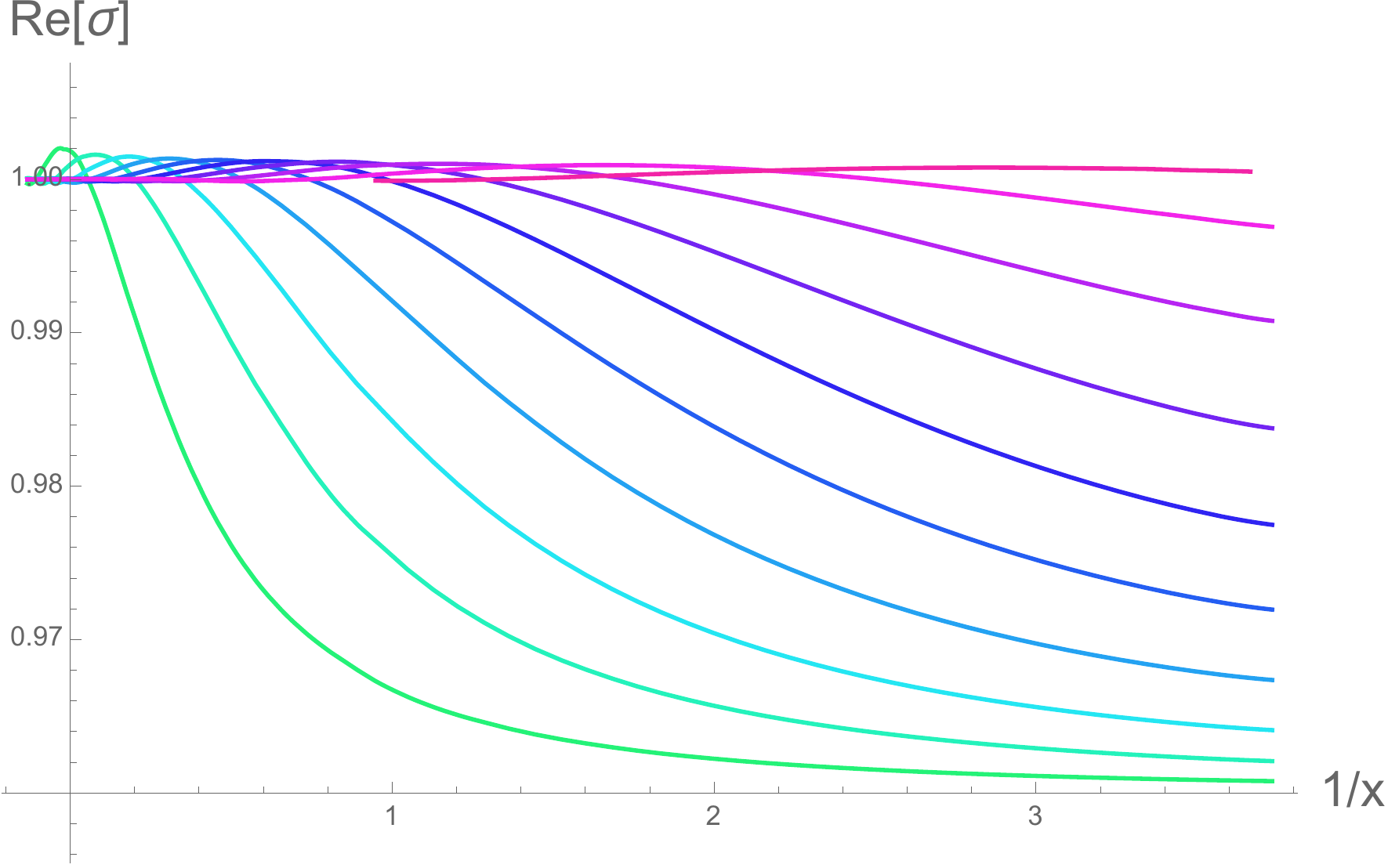}
\includegraphics[width=.06\textwidth]{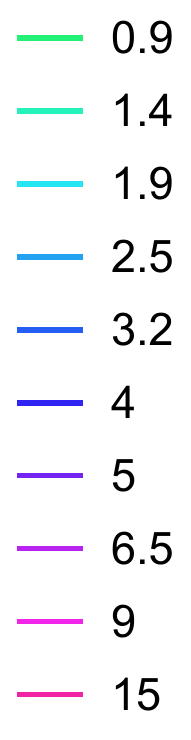}
\includegraphics[width=.44\textwidth]{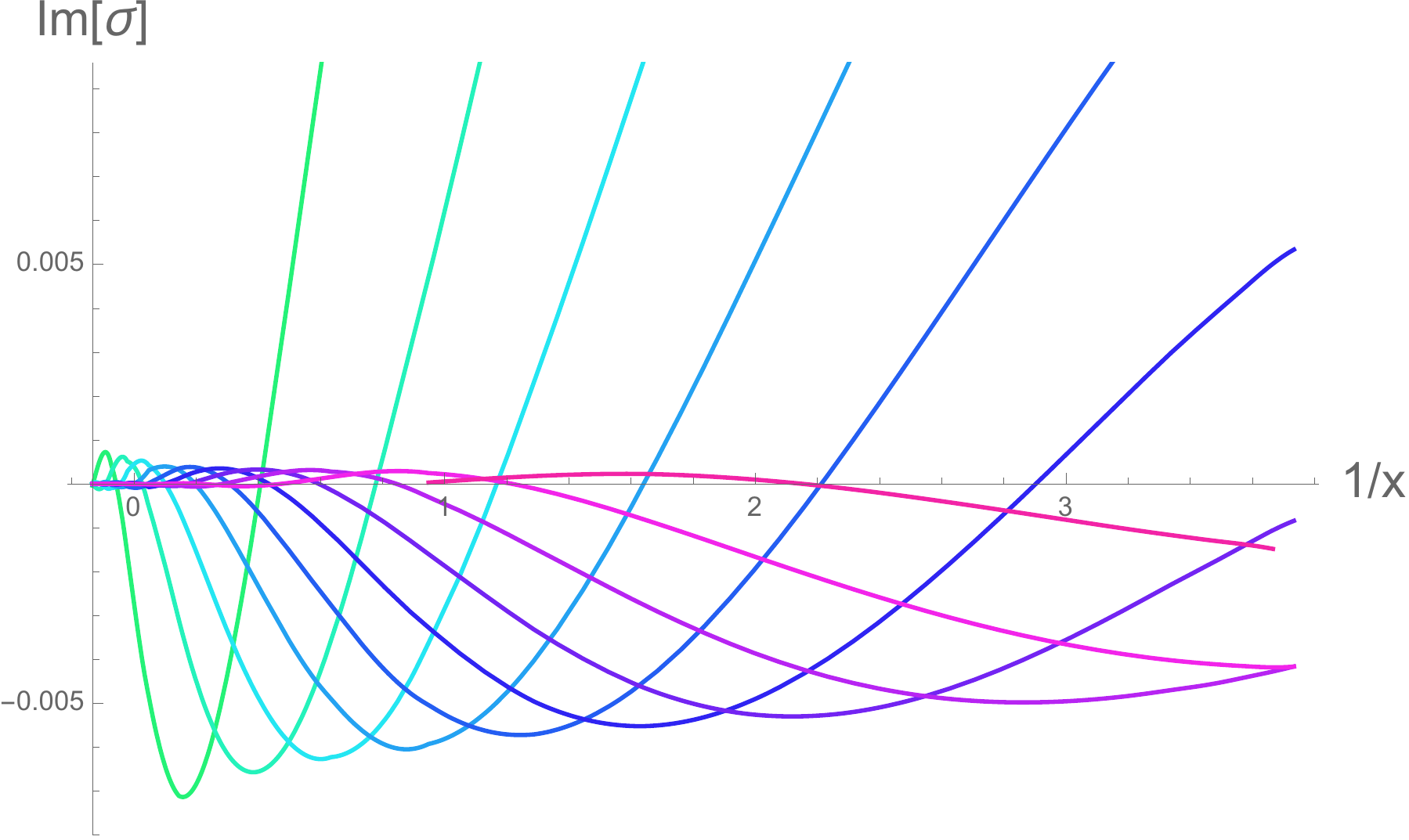}
\put(-212,107){\scriptsize $\omega/\mu$\normalsize}
\caption{\label{fig:figura1}{\underline{First line:}} real (left) and imaginary (right) parts of the conductivity $\sigma$ as a function of the ratio $\omega/\mu$ for different values of the doping. {\underline{Second line:}} real (left) and imaginary (right) parts of the conductivity $\sigma$ as a function of the doping, for the fixed values of $\omega/\mu$. The plots correspond to $m=0.15$ and $q_+=1$ with $q=0.2$ and $\mu^-=1.5$. 
}
\end{figure}

With this construction, the grand canonical potential in formula \eqref{eq:grand.canonical} can be plotted as a function of the doping ${\sf x}$. The resulting profiles for different values of the constants $q$ and $\mu^-$ are shown in Fig. \ref{fig:figura4}. Moreover, using  \eqref{eq:conductivitytilde} with constant $\sigma^-$, we map the conductivities $\sigma^+$ into that of the charge carriers $\sigma$. Results are shown in the first line of Fig. \ref{fig:figura1}.

The vertical lines on the first line of Fig. \ref{fig:figura1} fix the values of $\omega/\mu$ for which the conductivity as a function of the doping can be plotted. The resulting plots are shown in the second line of Fig. \ref{fig:figura1}. The real part of the conductivity oscillates at negative or small positive values of $1/{\sf x}$, and the goes smoothly to a fixed asymptotic value independent of $\omega/\mu$. Such decay is slower for larger $\omega/\mu$. The imaginary part on the other hand, oscillates around zero with a decreasing amplitude as the doping decreases. The oscillations are faster for smaller $\omega/\mu$. 

\begin{figure}[ht]
\,
\includegraphics[width=.44\textwidth]{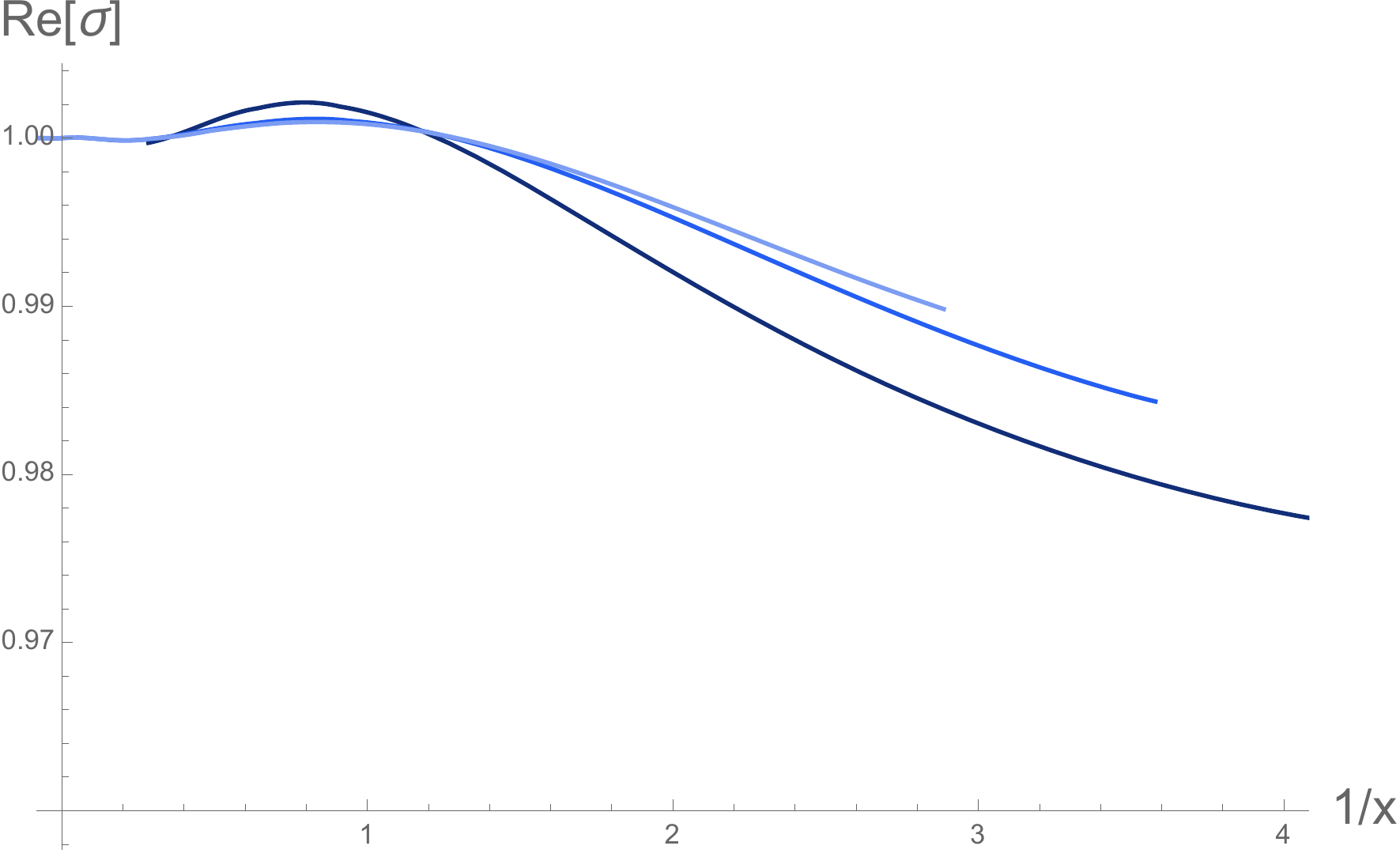}
\qquad\ \
\includegraphics[width=.44\textwidth]{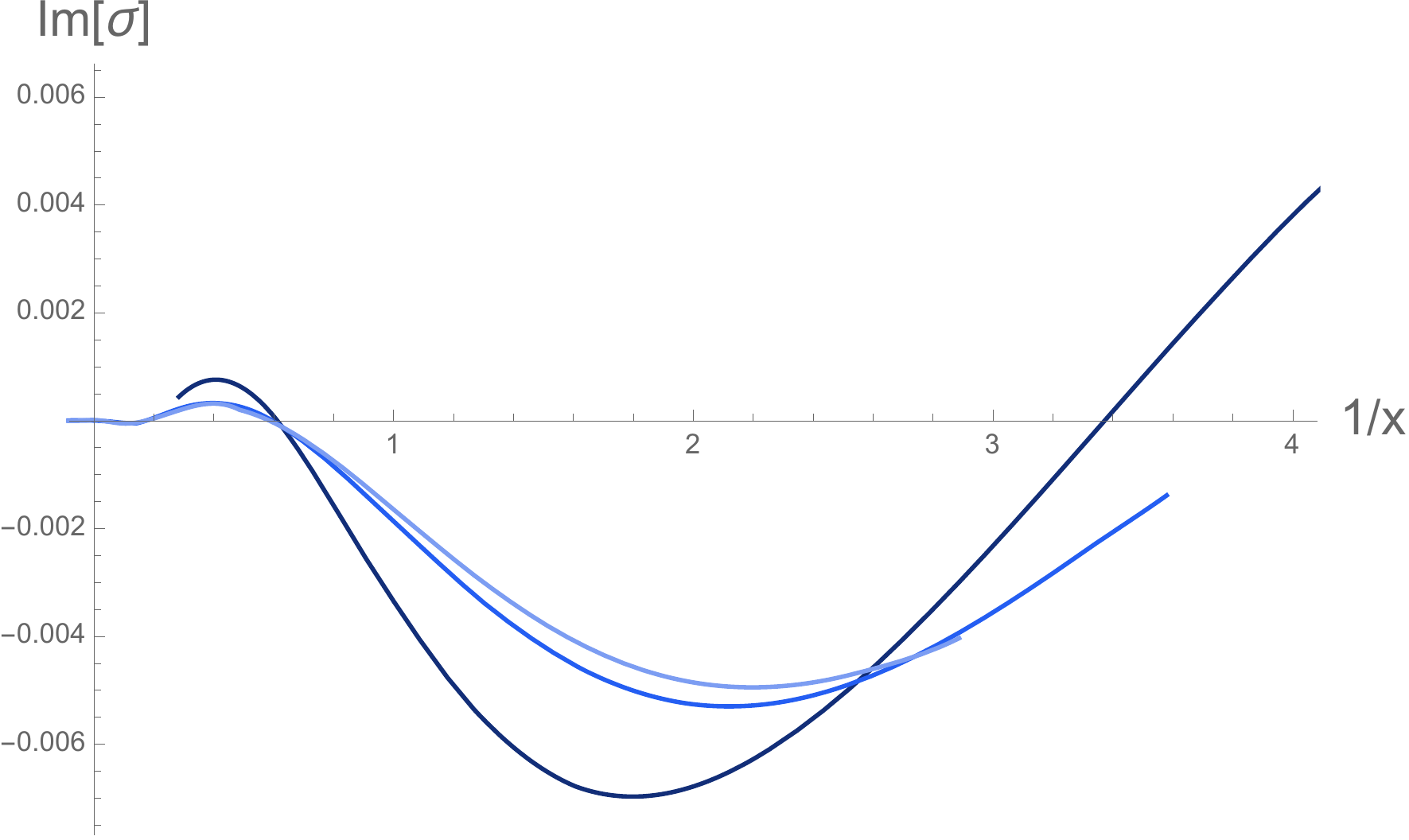}
\put(-278,73){\tiny $\mu^-\!=\!2.8$\normalsize}
\put(-251,60){\tiny $\mu^-\!=\!1.5$\normalsize}
\put(-231,45){\tiny $\mu^-\!=\!0.2$\normalsize}
\put(-10,87){\tiny $\mu^-\!=\!0.2$\normalsize}
\put(-30,47){\tiny $\mu^-\!=\!1.5$\normalsize}
\put(-57,25){\tiny $\mu^-\!=\!2.8$\normalsize}
\caption{\label{fig:figura2} Dependence on the parameter $\mu^-$ of the real (left) and imaginary (right) parts of the conductivity as a function of the doping. Here $m=0.15$ and $q_+=1$ with $q=0.2$. }
\end{figure}

\begin{figure}[ht]
\,
\includegraphics[width=.44\textwidth]{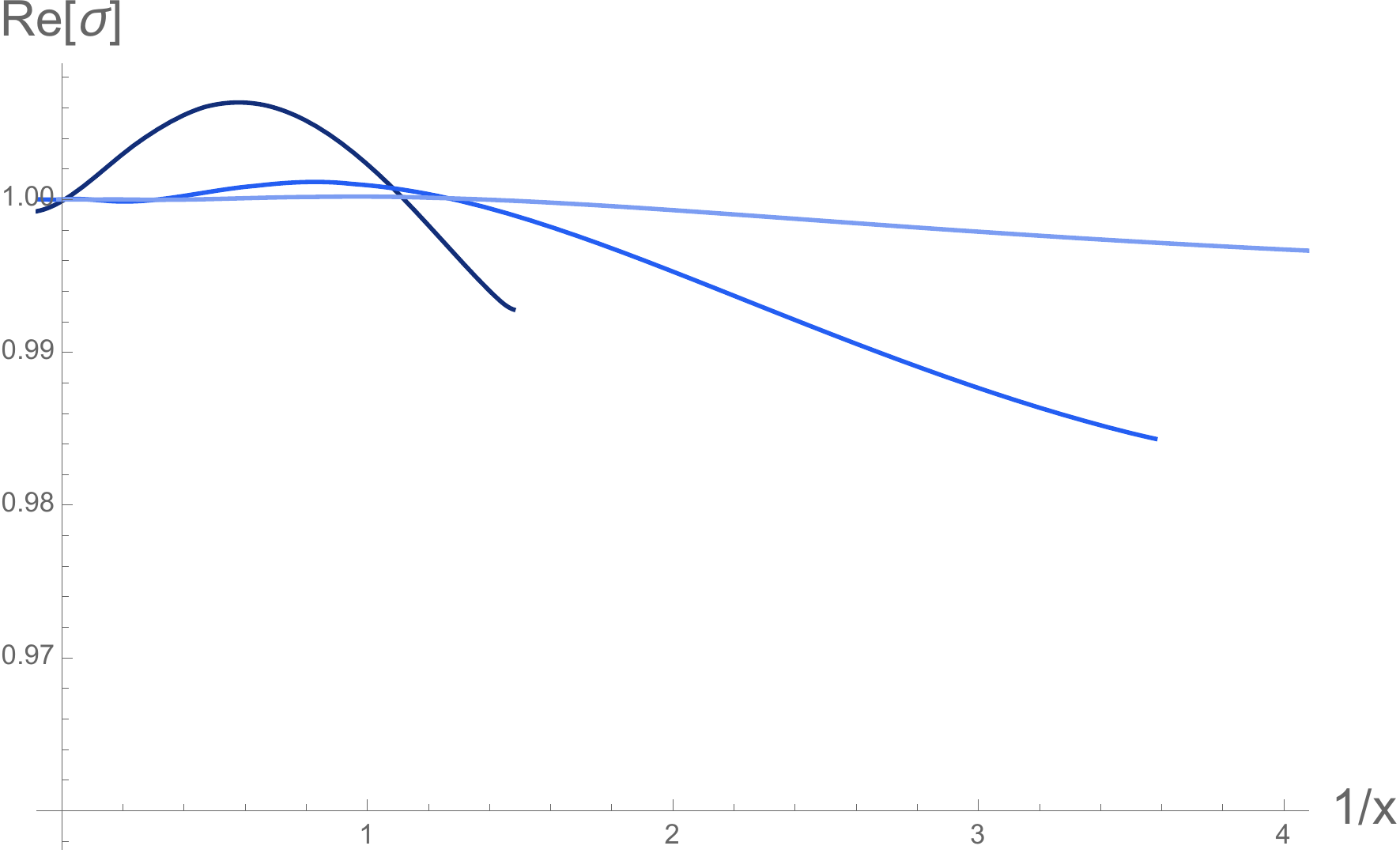}
\qquad\ \
\includegraphics[width=.44\textwidth]{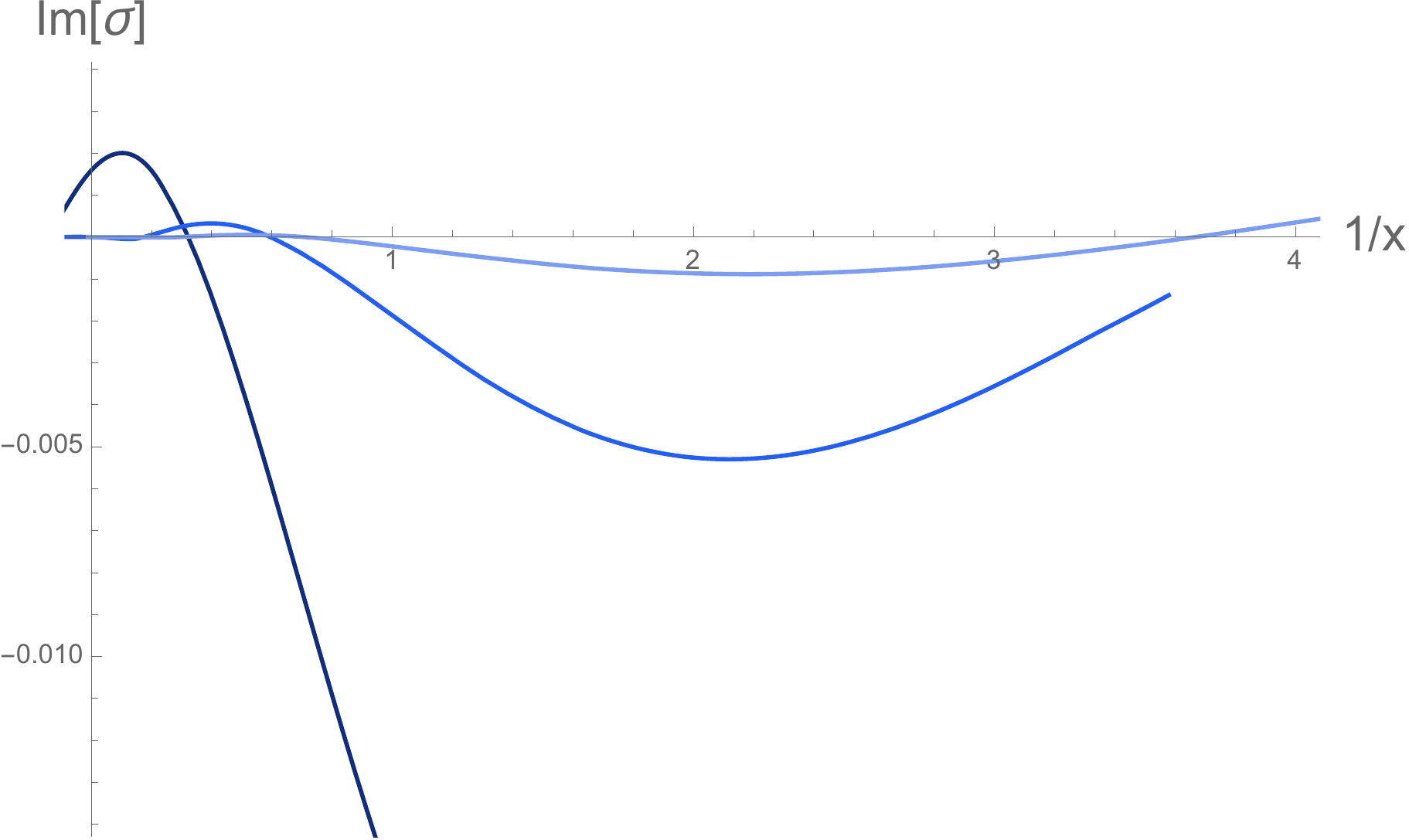}
\put(-232,81){\tiny $q\!=\!0.08$\normalsize}
\put(-252,55){\tiny $q\!=\!0.2$\normalsize}
\put(-340,68){\tiny $q\!=\!0.5$\normalsize}
\put(-15,88){\tiny $q\!=\!0.08$\normalsize}
\put(-30,72){\tiny $q\!=\!0.2$\normalsize}
\put(-137,0){\tiny $q\!=\!0.5$\normalsize}
\caption{\label{fig:figura3} Dependence on the parameter $q$ of the real (left) and imaginary (right) parts of the  conductivity as a function of the doping. Here $m=0.15$ and $q_+=1$ with $\mu^-=1.5$.}
\end{figure}

Changing the value of $\mu^-$ scales the plots on the first line of Fig. \ref{fig:figura1} horizontally, resulting in a relabelling of the positions of the vertical lines at fixed $\omega/\mu$. In consequence, the effect on the plots of the second line is to relabel the value of $\omega/\mu$ corresponding to each curve, see Fig. \ref{fig:figura2}. On the other hand, changing the value of $q$ scales the plots on the first line vertically, since it affects the values of $\sigma_0$ and $\sigma_1$. Then in the plots on the second line, it changes the range in which the conductivities take their values, see Fig \ref{fig:figura3}. 

~

As a conclusion, the doped holographic metal is a poorer conductor for small values of the doping, becoming a better one as the doping grows. The range and speed of the variation of the conductivity depends on the parameters of the mapping $q$ and $\mu^-$. There is an optimal doping at which the conductivity has a maximum.

\acknowledgments
The authors thank Guillermo Silva, Diego Correa, Gast\'on Giordano and Ignacio Salazar-Landea for useful comments. This work was funded by the CONICET grants PIP-2017-1109 and PUE 084 ``B\'usqueda de Nueva F\'\i sica'' and UNLP grant PID-X791.

\vspace{1.2cm}

\newpage 

~ 

\newpage

\bibliographystyle{JHEP}
\bibliography{references.bib}

\end{document}